\documentclass[aps,prd,preprint,groupedaddress]{revtex4-1}

\usepackage{graphicx} 
\usepackage{xcolor}
\begin{document}

\title{ Quasinormal modes for quarkonium in a plasma with magnetic fields     }

\author{Nelson R. F. Braga}\email{braga@if.ufrj.br}
\affiliation{Instituto de F\'{\i}sica,
Universidade Federal do Rio de Janeiro, Caixa Postal 68528, RJ
21941-972 -- Brazil}
 
\author{Luiz F.  Ferreira  }\email{luiz.faulhaber@ufabc.edu.br }
\affiliation{CCNH, Universidade Federal do ABC – UFABC, 09210-580, Santo Andr\'e -- Brazil.}


\begin{abstract} 
Heavy vector mesons detected after a  heavy ion collision are  important sources of  information about the  quark gluon plasma. The fraction of such particles that survive the plasma phase and reach the detectors  is related to the dissociation degree inside the thermal medium. 
A consistent picture for the thermal behavior of charmonium and bottomonium quasi-states  in a thermal medium was obtained recently using a holographic bottom up model.
This model captures the heavy flavour spectroscopy of masses and decay constants in the vacuum (zero temperature)  and is  consistently extended to finite temperature.  The   spectral functions that emerge provide  a description of the dissociation process in terms of the broadening  of the quasi-state peaks with temperature.  
 The holographic approach makes it possible to determine also the quasinormal modes. They are gravity solutions representing the  quasi-particle states in the thermal medium, with complex frequencies  related to the thermal mass and width. The quasinormal modes for charmonium and bottomonium have been studied very recently and a consistent description of the dissociation process was found. 
 
 An additional factor can  affect the dissociation process:  strong magnetic fields are expected to be present when the plasma is formed by non-central heavy ion collisions. So, it is important to understand the effect of  such fields on   the heavy meson dissociation scenario.  Here we extend the holographic determination of quasinormal modes for the case when  magnetic fields are present.   
The real and imaginary parts of the mode frequencies are determined for different values of  background $e B$ field. The associated dispersion relations for heavy quarks moving inside the plasma are  also investigated for both  $ c  \bar c$ and $ b \bar b  \, $.  

\end{abstract}


\maketitle

 \section{ Introduction }

 Heavy ion collisions, produced in particle accelerators, lead to the formation of a special type of matter: the quark gluon plasma. During the very short period of time when this plasma survives,  gluons and  light quarks live in a deconfined phase that behaves as a perfect fluid in approximate thermal equlibrium. All the information we have about this plasma phase is obtained indirectly, through  the analysis of the final products of heavy ion collisions that reach the detectors. 
The thermal behavior of heavy vector mesons, formed by   $b \bar b$ (bottomonium)  and $ c \bar c $ (charmonium), quark anti-quark pairs is quite different from the case of light mesons. For the dominant light flavours,  the dissociation occurs at a  (deconfinement) temperature when the thermal medium is formed.  On the other hand, heavy mesons undergo dissociation at higher temperatures. Depending on the temperature reached by the plasma,  they may survive the plasma phase or undergo  just a partial dissociation.  This is the reason why the fraction of  heavy mesons observed after a heavy ion collision
can  provide information about the  quark gluon plasma\cite{Matsui:1986dk,Satz:2005hx}. 

In order to translate information about the fraction of heavy mesons produced in a collision into knowledge about properties of the plasma, like temperature, density (chemical potential) and background magnetic field, one needs to understand how the dissociation process is influenced by these factors. 
 An important tool to model the thermal behavior of heavy vector mesons is the use of holographic bottom up models.
 A consistent description of the dissociation process using a holographic model containing three energy parameters was obtained in Refs. \cite{Braga:2017bml,Braga:2018zlu,Braga:2018hjt}. 
 These parameters have a simple physical interpretation. One represents the quark mass, another is related to the string tension and the third one is an  ultraviolet (UV) energy  parameter, necessary in order  to fit the  decay constant spectra.  This UV  parameter is related to the large mass change that occurs in a non hadronic decay, when a heavy vector meson  transforms into  light leptons. 
 This model is an improvement of previous approaches presented in refs. \cite{Braga:2015jca,Braga:2016wkm,Braga:2017oqw} that used just two energy parameters. 
 
 The thermal behavior of heavy vector mesons in a plasma was described  in  \cite{Braga:2017bml} in terms of the spectral function, that exhibits a clear picture of the dissociation as the broadening of the quasi-particle peaks. Then in \cite{Braga:2018zlu} the determination of the spectral functions  was extended to the case when a magnetic background field is present. 
 
 An alternative way of describing the dissociation process is analysing the spectra of quasinormal modes (QNM). In the holographic description of a gauge field theory, the QNM are gravitational field solutions that represent the quasi-particle states. They are solutions of the equations of motion that satisfy an incoming wave condition on the horizon and vanish on the boundary. The corresponding frequencies are complex.  The real part represents the thermal mass and the imaginary part ir relates with the thermal width of the quasi-state. The calculation of the QNM spectra for charmonium and bottomonium was presented in  \cite{Braga:2018hjt}.
  
The dissociation process is affected when magnetic fields are present in the plasma. This situation is important because   strong magnetic fields can be produced in non central heavy ion collisions because of the motion of the electric charges \cite{Kharzeev:2007jp,Fukushima:2008xe,Skokov:2009qp}. The presence of a magnetic field $e B$ has direct consequences for the plasma.  A decrease in the QCD deconfinement temperature with increasing field was observed in  Lattice calculations \cite{Bali:2011qj}, using the  MIT bag model\cite{Fraga:2012fs} and using the holographic D4-D8 model\cite{Ballon-Bayona:2013cta}. The effect of a magnetic field in the transition temperature of a plasma has been studied using holographic models in many works, as for example \cite{Mamo:2015dea,Dudal:2015wfn,Evans:2016jzo,Li:2016gfn,Ballon-Bayona:2017dvv,Rodrigues:2017cha}.  Other studies of the effect of magnetic field in quarkonia or in baryons can be found, for example, in \cite{Iwasaki:2018pby,Giataganas:2018uuw,Iwasaki:2018czv,Bonati:2018uwh}.

The presence of magnetic field can also affect the behavior of heavy mesons \cite{Braga:2018zlu}, changing the dissociation temperature. Here we will investigate this issue using quasinormal modes.  We will extend the  calculation of quasinormal modes presented in  \cite{Braga:2018hjt}  to the case when a magnetic field is present.    The dependence of the real and imaginary part of the frequency on the magnetic field will be investigated for charmonium and bottomonium states. Then the dispersion relations will be considered. The dependence of the complex frequencies on the linear momentum, for hadrons moving inside a plasma with a magnetic field will be analysed. 
This type of study provides a detailed picture of the thermal behavior of the heavy vector mesons in
a medium like the quark gluon plasma. 

The article is organized as follows.  In section { \bf 2} we review the holographic AdS/QCD model for heavy vector mesons. Then  in  section {\bf 3} we study the spectral function for heavy quarks in motion inside a plasma with a magnetic field. The quasinormal mode frequencies are investigated in section {\bf 4} and some final comments are left for section {\bf 5}. 
 \bigskip 

\section{AdS/QCD description of heavy vector mesons }
 
 As discussed in refs.\cite{Braga:2017bml,Braga:2018zlu,Braga:2018hjt}, a   consistent description of heavy vector mesons in a thermal medium is obtained by
 considering a  5-d anti-de Sitter black hole space-time
 \begin{equation}
 ds^2 \,\,= \,\, \frac{R^2}{z^2}  \,  \Big(  -  f(z) dt^2 + d\vec{x}\cdot d\vec{x}  + \frac{dz^2}{f(z) }    \Big)   \,,
 \label{metric2}
\end{equation}
with $ f (z) = 1 - \frac{z^ 4}{z_h^4} $, where $z_h$ is the horizon position. The condition of regularity of the metric,  in the imaginary time formulation where the time variable is periodic with period $ 0 \le t \le \beta = 1/T$,   leads to the following relation between the horizon position and the temperature:
\begin{equation} 
T =  \frac{\vert  f'(z)\vert_{(z=z_h)}}{4 \pi  } = \frac{1}{\pi z_h}\,.
\label{temp}
\end{equation}
 
 The holographic  dual of the heavy  mesons are vector fields $V_m = (V_\mu,V_z)\,$ ($\mu = 0,1,2,3$) living in the space (\ref{metric2}). One can gauge away the $z$ component: $V_z=0$ and 
  take the  $V_\mu $ components as sources for the correlators of the gauge theory currents $ J^\mu = \bar{\psi}\gamma^\mu \psi \,$. 
  
The action integral for the vector field is:
\begin{equation}
I \,=\, \int d^4x dz \, \sqrt{-g} \,\, e^{- \phi (z)  } \, \left\{  - \frac{1}{4 g_5^2} F_{mn} F^{mn}
\,  \right\} \,\,, 
\label{vectorfieldaction}
\end{equation}
where $F_{mn} = \partial_m V_n - \partial_n V_m$. The energy parameters of the model are introduced through the background scalar field $\phi(z)$ with the form:  
\begin{equation}
\phi(z)=k^2z^2+Mz+\tanh\left(\frac{1}{Mz}-\frac{k}{ \sqrt{\Gamma}}\right)\,.
\label{dilatonModi}
\end{equation} 

 The parameters have the following physical interpretation:  $k$  is related to the  quark mass, 
  $\Gamma $  is related to the string tension of the quark anti-quark interaction. The  parameter $M$ 
 represents effectively the mass scale of a non hadronic decay, characterized by a matrix element of the form   $ \langle 0 \vert \, J_\mu (0)  \,  \vert n \rangle = \epsilon_\mu f_n m_n \,, $
representing a transition from a meson  at radial excitation level $n$ to the hadronic vacuum. 
 The decay constant $ f_n $ is essentially proportional to this matrix element and  the large mass parameter $M$ makes it possible to fit the corresponding spectra.  
 
 In the zero temperature case $ f (z) = 1  $ and the equation of motion coming from action (\ref{vectorfieldaction}) for the $ \mu = (1,2,3)$ components of the vector field, denoted here generically as $V$, in momentum space reads
\begin{equation}
\partial_{z} \left[  \frac{R}{z} e^{-\phi(z)} \partial_{z} V (p,z)  \right]-p^2  \frac{R}{z} e^{-\phi(z)} V (p,z)  = 0\,.
\label{eqmotion}
\end{equation}
 The normalizable solutions, that represent vector meson states, are the ones that satisfy the  
boundary condition  $ V (p, z=0) =0$. They form a discrete spectrum  of $p^2=-m_{n}^{2}$ where $m_n$ are interpreted as  the masses of    meson states at the $n^{th}$ excitation level. 
 
Decay constants are  proportional to the transition matrix from  the vector meson $n$  state to the vacuum and are given by\cite{Braga:2017bml} 
\begin{equation}
f_n=\frac{1}{g_{5} m_{n}}\lim\limits_{z \rightarrow 0} \left( \frac{R}{z} e^{-\phi(z)} \partial_z  \Psi_n(z)\right) \,.
\label{decayconstant}
\end{equation}

 The  values of the parameters that  provide a fit to charmonium and bottomonium spectra are respectively:
\begin{eqnarray}
  k_c  = 1.2  \, {\rm GeV } ; \,\,   \sqrt{\Gamma_c } = 0.55  \, {\rm GeV } ; \,\, M_c=2.2  \, {\rm GeV }\,;
  \label{parameters1}
  \\
   k_b = 2.45  \, {\rm GeV } ; \,\,   \sqrt{\Gamma_b } = 1.55  \, {\rm GeV } ; \,\, M_b=6.2  \, {\rm GeV }\,.
  \label{parameters2}
  \end{eqnarray}  
\begin{table}[h]
\centering
\begin{tabular}[c]{|c||c||c|}
\hline 
\multicolumn{3}{|c|}{  Holographic (and experimental) results for bottomonium   } \\
\hline
 State &  Mass (MeV)     &   Decay constants (MeV) \\
\hline
$\,\,\,\, 1S \,\,\,\,$ & $ 6905 \,\,(9460.30\pm 0.26) $  & $ 719 \,( 715.0 \pm 2.4) $ \\
\hline
$\,\,\,\, 2S \,\,\,\,$ & $   8871 \,( 10023.26 \pm 0.32) $   & $ 521 \,(497.4 \pm 2.2) $  \\
\hline 
$\,\,\,\,3S \,\,\,\,$ & $  10442 \, \,( 10355.2 \pm 0.5) $   & $427 \, (430.1  \pm 1.9) $ \\ 
\hline
$ \,\,\,\, 4S  \,\,\,\,$ & $ 11772 \, (10579.4 \pm 1.2)  $  & $ 375 \,(340.7  \pm 9.1)$ \\
\hline
\end{tabular}   
\caption{Holographic masses and decay constants for the bottomonium $S$-wave resonances. Experimental values inside parentheses for comparison. }
\end{table}

   \begin{table}[h]
\centering
\begin{tabular}[c]{|c||c||c|}
\hline 
\multicolumn{3}{|c|}{  Holographic (and experimental)  results for charmonium   } \\
\hline
 State &  Mass (MeV)     &   Decay constants (MeV) \\
\hline
$\,\,\,\, 1$S$ \,\,\,\,$ & $ 2943 \,\, (3096.916\pm 0.011)  $  & $ 399 \, (416 \pm 5.3)$ \\
\hline
$\,\,\,\, 2$S$ \,\,\,\,$ & $  3959 \,\, (3686.109 \pm 0.012) $   & $ 255  \, (296.1 \pm 2.5)$  \\
\hline 
$\,\,\,\,3$S$ \,\,\,\,$ & $  4757 \,\, (4039 \pm 1 ) $   & $198 \, ( 187.1  \pm 7.6) $ \\ 
\hline
$ \,\,\,\, 4$S$  \,\,\,\,$ & $ 5426\,\,  (4421 \pm 4)  $  & $ 169 \,  (160.8  \pm 9.7)$ \\
\hline
\end{tabular}   
\caption{Holographic masses and decay constants for the charmonium $S$-wave resonances. Experimental values inside parentheses for comparison.  }
\end{table}

 These parameters lead to the masses and decay constants calculated in in ref.  \cite{Braga:2018zlu} , that we show on Tables 1 and 2. For comparison we show inside parentheses the corresponding available experimental  data.  

It is worth to remark that in order to find an  appropriate description of the finite temperature  behavior of  the heavy mesons, it is necessary to find a nice fit of the decay constants. 
The reason is that the thermal spectral function is the imaginary part of the retarded Green's function. 
The relevant part of the Green's function is the two point function that, at zero temperature, has the following  spectral decomposition  in terms of masses $m_n$ and   decay constants $f_n$:  
 \begin{equation}
\Pi (p^2)  = \sum_{n=1}^\infty \, \frac{f_n^ 2}{(- p^ 2) - m_n^ 2 + i \epsilon} \,.
\label{2point}
\end{equation}
The imaginary part of eq.(\ref{2point}) is a sum of delta functions 
$$ f_n^2 \, \delta ( - p^2 - m_n^2 )\,, $$ 
\noindent with coefficients proportional to  the square of the decay constants. 
So, the extension to finite temperature, when the  quasi-particle state
 peaks have a finite width, must be consistent with the zero temperature behavior of the decay constants. 
 The original bottom up holographic models, like the precursor hard wall model \cite{Polchinski:2001tt,BoschiFilho:2002ta,BoschiFilho:2002vd}, provide decay constants that behave inconsistently with the experimental results.  The values of $f_n$  for the hard wall increase with   radial level $n$, the opposite of the behavior observed from experiments, and shown on tables  1 and 2. 
 
 The model revised in this section was applied in refs \cite{Braga:2017bml,Braga:2018zlu} to calculate the spectral functions with and without magnetic fields and then in \cite{Braga:2018hjt} to find out the quasinormal modes for a thermal medium without magnetic field.


\section{ Spectral function for a moving heavy meson in the presence of a magnetic field}

\subsection{ Dual background}

In order to find a geometry that represents a thermal medium with a background magnetic field, one starts with the Einstein-Maxwell action is given by:
\begin{equation}\label{bulkaction}
S=\frac{1}{16 \pi G_{5}}\int d^{5}x \sqrt{-g}\left(R-F^{MN}F_{MN}+\frac{12}{L^2} \right)+S_{GH}
\end{equation}
where $R$ is the Ricci scalar, the factor $\, \frac{12}{L^2}$ is minus the cosmological constant,  $F_{MN}$ is the electromagnetic field strength and $ S_{GH} $ is a surface term.  
  
Varying the action (\ref{bulkaction}) with respect to the fields, one obtains 
\begin{eqnarray}
R_{MN}=&-&\frac{4}{L^2}g_{MN}-\frac{g_{MN}}{3}F^{PQ}F_{PQ}
+ 2F_{MP}F_{\,\,\,N}^{P},
\end{eqnarray}
\begin{eqnarray}
\nabla_{M}F_{MN}=0.
\end{eqnarray}
 The holographic dual of a medium where there is a constant magnetic field $eB$ was presented in   \cite{Dudal:2015wfn} and reads
\begin{equation}
 ds^2 \,\,= \,\, \frac{R^2}{z^2}  \,  \Big(  -  f(z) dt^2 + \frac{dz^2}{f(z) }  +( dx_1^2+dx_2^2)d(z)+dx_3^2h(z)  \Big)   \,.
 \label{metric3}
\end{equation}
The metric factors depend on the horizon position and on the magnetic field $ eB$, that points in the  $x_3 $ direction
\begin{equation}
f (z) = 1 - \frac{z^ 4}{z_h^4}+\frac{2}{3}\frac{e^2B^2z^4}{1.6^2}\ln\left(\frac{z}{z_h}\right)\, ,
\end{equation}
\begin{equation}
h(z) = 1 + \frac{8}{3}\frac{e^2B^2}{1.6^2}\int^{1/z}_{+\infty}dx\frac{\ln{(z_hz)}}{z^3(z^2-\frac{1}{z_h^4x^2})}.
\end{equation}
\begin{equation}
d(z) = 1 - \frac{4}{3}\frac{e^2B^2}{1.6^2}\int^{1/z}_{+\infty}dx\frac{\ln{(z_hz)}}{z^3(z^2-\frac{1}{z_h^4x^2})}\, .
\end{equation}
 
The temperature is: 
\begin{equation} 
T =  \frac{\vert  f'(z)\vert_{(z=z_h)}}{4 \pi  } = \frac{1}{4\pi}\left\vert \frac{4}{z_h}-\frac{2}{3}e^2B^2z_{h}^3\right\vert \,.
\label{temp}
\end{equation}


\subsection{Equations of motion}

We consider  that the  action for the vector field, that represents vector mesons,   has the same form  of eq. (\ref{vectorfieldaction}),  with the dilaton background $ \phi (z) $ of 
eq. (\ref{dilatonModi}) but with the metric of eq.  (\ref{metric3}).   
In addition, we choose the radial gauge $ V_{z} =0 $ and consider plane wave solutions  propagating in the magnetic field  direction, $x_{3}$: $V_{\mu} = e^{-i\omega t+i q x_{3}}V_{\mu}(z,\omega,q)$ with the wave vector $p_{\mu}=(-\omega,0,0,q)$. The equations of motion are given by
\begin{equation}\label{eqtt}
V_{t}''-\left(\frac{1}{z}+\phi'-\frac{d'}{d}-\frac{h'}{2h} \right)V_{t}'-\frac{q}{f h}\left(qV_{t}+\omega V_{3}\right)=0,
\end{equation}

\begin{equation}\label{eqx3x3}
V_{3}''+\left(\frac{f'}{f}-\frac{1}{z}-\phi'+\frac{d'}{d}-\frac{h'}{2h} \right)V_{3}'+\frac{\omega}{f^2}\left(qV_{t}+\omega V_{3}\right)=0, \,\,\,\,\,\,\,\,\,\,\ 
\end{equation}

\begin{equation}\label{eqx1x1}
V_{\beta}''+\left(\frac{f'}{f}-\frac{1}{z}-\phi'+\frac{h'}{2h} \right)V_{\beta}'+\frac{1}{f^2}\left(\omega^2-q^2\frac{f}{h} \right)V_{\beta}=0 \, , \,\,\,\,\,\,\,\,\,\,\ (\beta=1,2)
\end{equation}

\begin{equation}\label{eqzz}
h\omega V_{t}'+qfV_{3}'=0
\end{equation} 

\noindent where  the  prime  ($'$)  denotes  the  derivative  with  respect  to z. One can write the previous equations in terms of the electric field components, $E_{1}=\omega V_{1}$, $E_{2}=\omega V_{2}$ and $E_{3}=\omega V_{3}+qV_{t}$, in the following form

\begin{equation}\label{eqTrans}
E_{\alpha}''+\left(\frac{f'}{f}-\frac{1}{z}-\phi'+\frac{h'}{2h} \right)E_{\alpha}'+\left(\frac{\omega^2}{f^2}-\frac{q^2}{fh}\right)E_{\alpha}=0 \, , \,\,\,\,\,\,\,\,\,\,\ (\alpha=1,2)
\end{equation} 
\\
\begin{equation}\label{eqLog}
E_{3}''+\left(\frac{f'}{f}\frac{\omega^2}{\omega^2 h-q^2f}-\frac{1}{z}-\phi' +\frac{d'}{d}-\frac{h'}{2h}\frac{\omega^2 h+q^2f}{\omega^2 h-q^2f} \right)E_{3}'+\left(\frac{\omega^2}{f^2}-\frac{q^2}{fh}\right)E_{3}=0. \,\,\,\,\,\,\,\,\,\,\ 
\end{equation}


\subsection{Spectral Function}

\begin{figure}[h]
\label{g67}
\begin{center}
\includegraphics[scale=0.35]{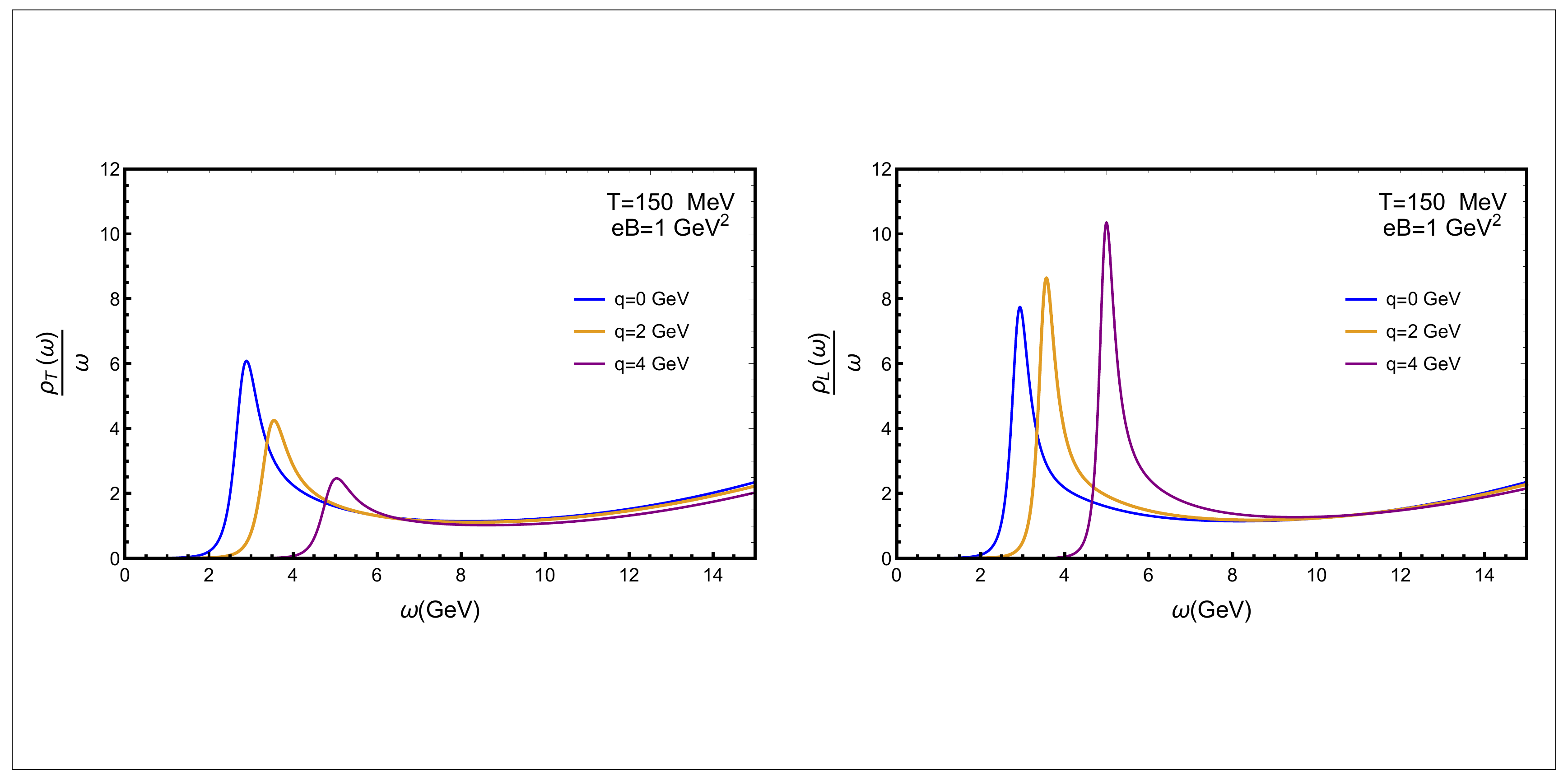}
\end{center}
\caption{   Charmonium spectral functions at T= 150 MeV,  eB = 1 GeV$^2$ for different values of linear momentum $q$. Transverse polarization on the left panel, longitudinal on the right panel }
\end{figure}

\begin{figure}[h]
\label{g67}
\begin{center}
\includegraphics[scale=0.35]{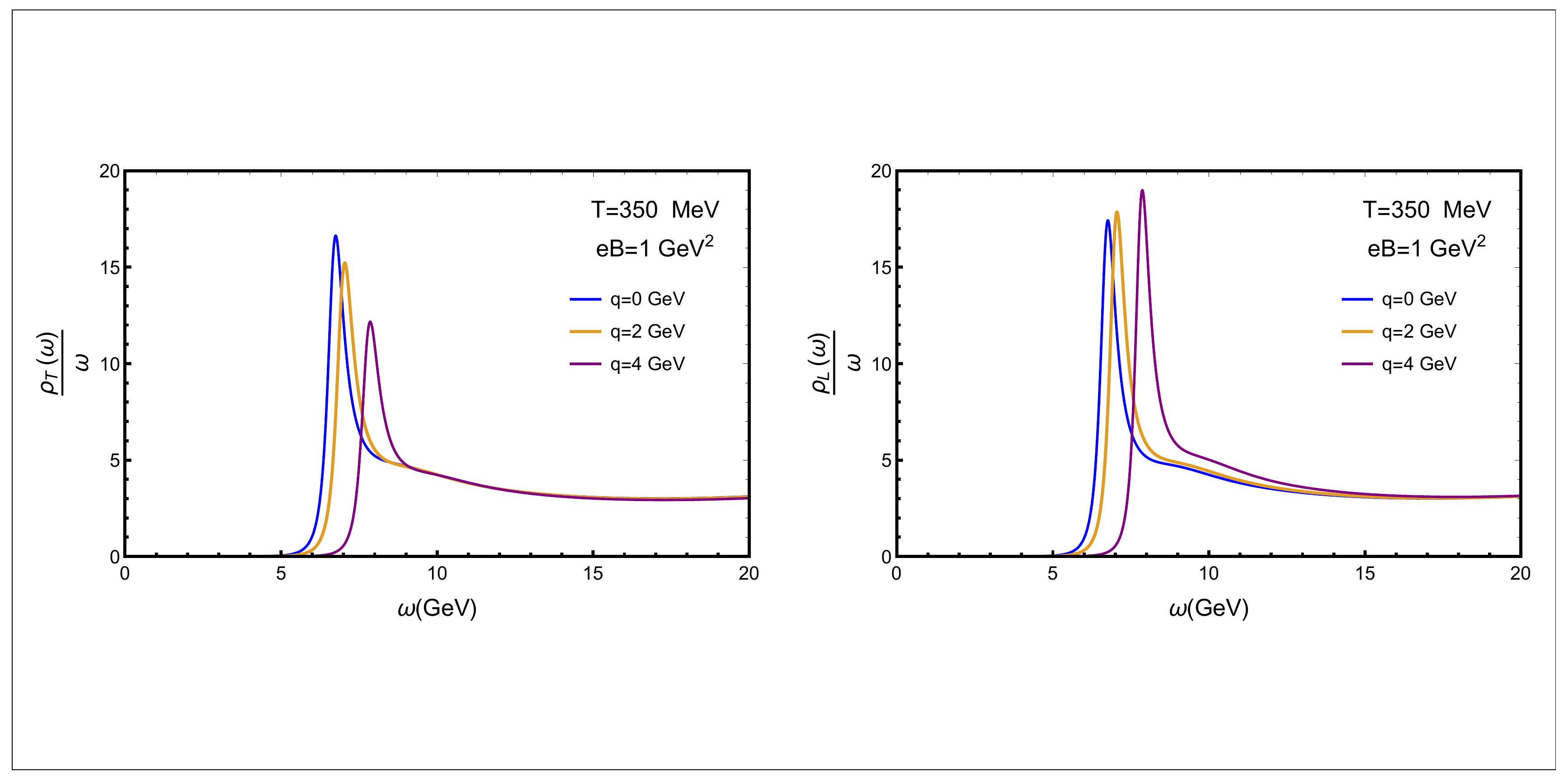}
\end{center}
\caption{  Bottomonium spectral functions at T= 350 MeV,  eB = 1 GeV$^2$ for different values of linear momentum $q$. Transverse polarization on the left panel, longitudinal on the right panel }
\end{figure}

 The spectral function is obtained from the retarded Green's functions of the vector meson currents \begin{equation}\label{Green}
G^{R}_{\mu \nu}(p)=-i\int d^4 x e^{- i p \cdot x}\theta(t)\langle \left[ J_{\mu}(x),J_{\nu}(0)\right]\rangle\,,
\end{equation}
 that can be obtained  following the  Son-Starinets prescription \cite{Son:2002sd}.  With  this purpose it is convenient to write the equations of motion (\ref{eqTrans}) and (\ref{eqLog}) in terms of the bulk to boundary propagator  $E_{j}(z,p)=\mathcal{E}_{j}(z,p)E^{0}_{j}(p)$, 
 \begin{equation}\label{eqTransx}
\mathcal{E}_{\alpha}''+\left(\frac{f'}{f}-\frac{1}{z}-\phi'+\frac{h'}{2h} \right)\mathcal{E}_{\alpha}'+\left(\frac{\omega^2}{f^2}-\frac{q^2}{fh}\right)\mathcal{E}_{\alpha}=0 \, , \,\,\,\,\,\,\,\,\,\,\ (\alpha=1,2)
\end{equation} 

\begin{equation}\label{eqLogx}
\mathcal{E}_{3}''+\left(\frac{f'}{f}\frac{\omega^2}{\omega^2 h-q^2f}-\frac{1}{z}-\phi' +\frac{d'}{d}-\frac{h'}{2h}\frac{\omega^2 h+q^2f}{\omega^2 h-q^2f} \right)\mathcal{E}_{3}'+\left(\frac{\omega^2}{f^2}-\frac{q^2}{fh}\right)\mathcal{E}_{3}=0. \,\,\,\,\,\,\,\,\,\,\ 
\end{equation}
and use the  relation established  by  the prescription \cite{Son:2002sd}:

\begin{eqnarray}\label{Greenrel}
&&\frac{G^{R}_{tt}}{q^2}=-\frac{G^{R}_{tx_{3}}}{q\omega}=-\frac{G^{R}_{3t}}{q\omega}=\frac{G^{R}_{33}}{\omega^2}=-\frac{R}{g^2_{5}}\frac{1}{\omega^2-q^2}\lim_{z \rightarrow 0} \frac{e^{-\phi}}{z}\partial_{z}\mathcal{E}_{3}(z,p)\cr \ && G^{R}_{11}=-\frac{R}{g^2_{5}}\lim_{z \rightarrow 0} \frac{e^{-\phi}}{z}\partial_{z}\mathcal{E}_{1}(z,p)\,, \,\,\, G^{R}_{22}=-\frac{R}{g^2_{5}}\lim_{z \rightarrow 0} \frac{e^{-\phi}}{z}\partial_{z}\mathcal{E}_{{2}}(z,p)\,.
\end{eqnarray}
The  longitudinal $G^{R}_{x_3 x_3}$ and transversal $G^{R}_{\alpha \alpha}$ Green's functions give us the 
 corresponding spectral functions 
\begin{equation}\label{eq1}
\rho_{33}(\omega,q)\equiv -2 ImG^{R}_{33}(\omega,q)=\frac{2R}{g^2_{5}}\frac{\omega^2}{\omega^2-q^2}\lim_{z \rightarrow 0} \frac{e^{-\phi}}{z}\partial_{z}\mathcal{E}_{{3}}(z,p)\,,
\end{equation} 
\begin{equation}\label{eq2}
\rho_{\alpha \alpha}(\omega,q)\equiv -2 ImG^{R}_{\alpha \alpha}(\omega,q)=-\frac{2R}{g^2_{5}}\lim_{z \rightarrow 0} \frac{e^{-\phi}}{z}\partial_{z}\mathcal{E}_{\alpha}(z,p)\,,
\end{equation}
where the bulk to boundary propagators $\mathcal{E}$ are the solutions of eqs. (\ref{eqTransx}) and (\ref{eqLogx}) satisfying the infalling  boundary condition
\begin{eqnarray}\label{boundary}
&& \mathcal{E}(z \rightarrow z_h,\omega) \longrightarrow \left(1-\frac{z}{z_h}\right)^{-i\omega/4\pi T}\,\left[1+a_1\left(1-\frac{z}{z_h}\right)+a_2\left(1-\frac{z}{z_h}\right)^2+...\right]\,,
\end{eqnarray}
and the bulk to boundary condition 
\begin{equation}
\label{boundary2}
\mathcal{E}(z\to 0,\omega)=1\,.
\end{equation}

Spectral functions for heavy vector mesons at rest  inside a plasma with a magnetic field were studied already in ref. \cite{Braga:2018zlu}.  Here we extend this study to the case when mesons are in motion relative to the medium with magnetic field in order to see the influence of the state of motion in  the dissociation process. With this purpose we solved equations (\ref{eqTransx}) and (\ref{eqLogx}) for different values of momentum $q$,  with $eB$ = 1 GeV and $ T= 150 $ GeV for charmonium and then $T= 350$ GeV for bottomonium. 
The results are show in  Figures {\bf 1} and {\bf 2}. On the left panels we show the cases when the motion is in the same direction of the magnetic field but perpendicular to the polarization. One notes clearly that the dissociation degree increases with the linear momentum. The faster the meson moves, the broader the peak  becomes. The opposite result shows up on the right panels, where the results for polarization in the same direction of the motion and the field are ploted. In this case the peaks get higher as the meson moves faster, indicating that the dissociation degree decreases with linear momentum.

\section{Quasinormal modes}  
 
 \subsection{Physical  meaning }

\begin{figure}[h]
\label{g67}
\begin{center}
\includegraphics[scale=0.35]{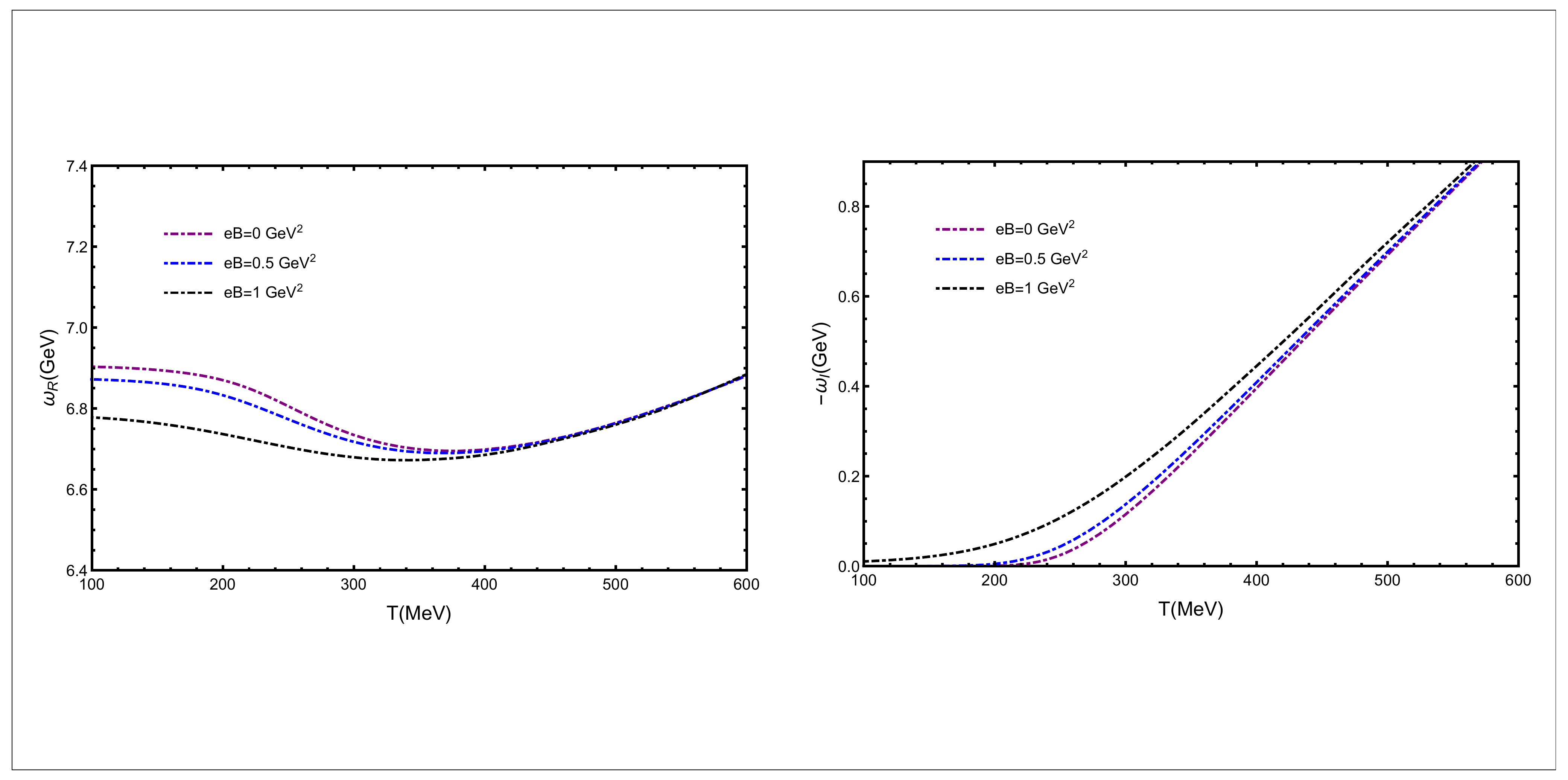}
\end{center}
\caption{Real (left panel) and imaginary (right panel) parts of the quasinormal mode frequency of bottomonium with transverse polarization as a function of the temperature for different values of magnetic field.     }
\end{figure}

\begin{figure}[h]
\label{g67}
\begin{center}
\includegraphics[scale=0.35]{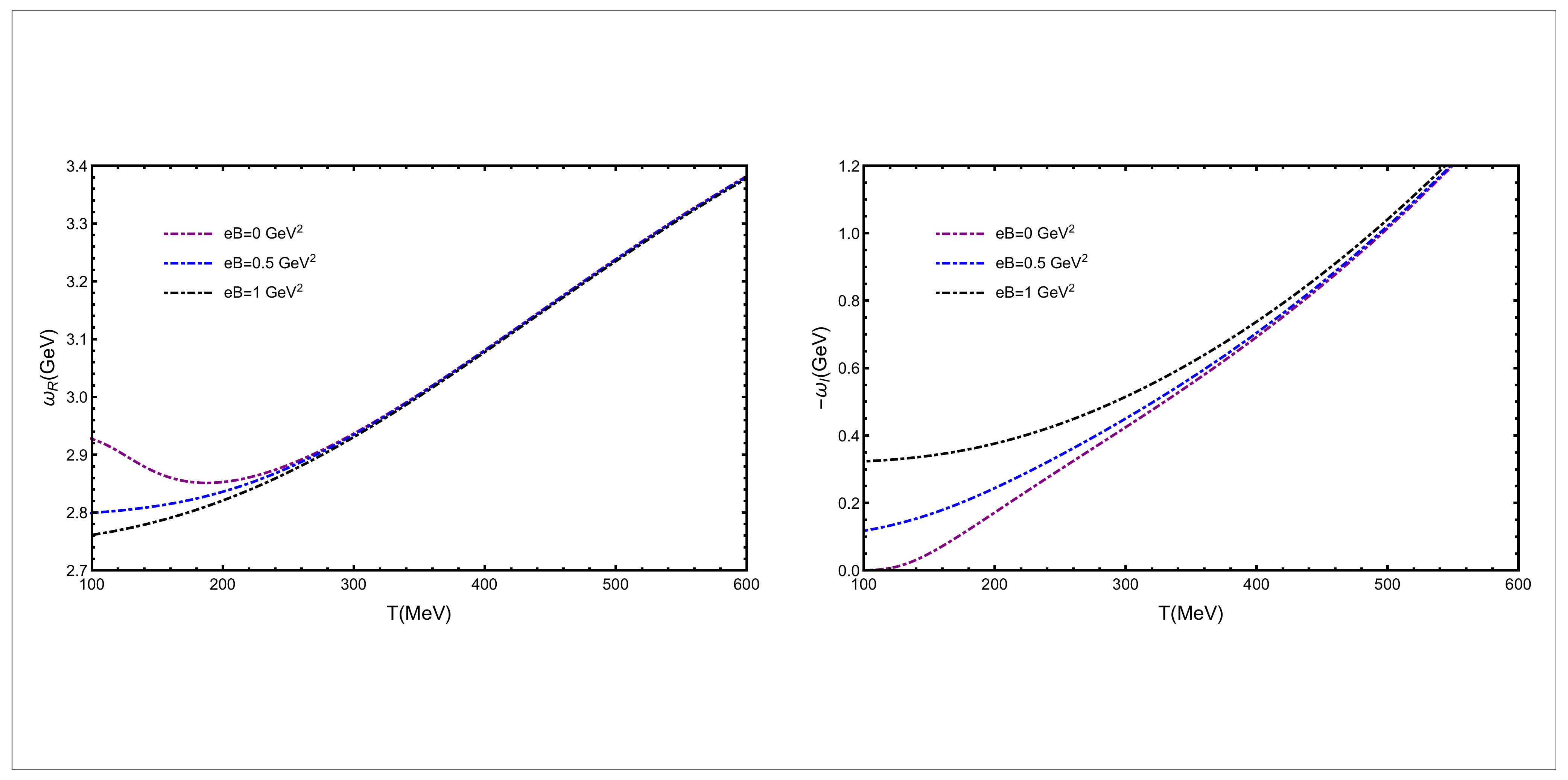}
\end{center}
\caption{Real (left panel) and imaginary (right panel) parts of the quasinormal mode frequency of charmonium with transverse polarization as a function of the temperature for different values of magnetic field.     }
\end{figure}

 In gauge/string duality at zero temperature, hadronic states are represented by normalizable solutions of gravity fields. These solutions, also called normal modes,  are found by imposing boundary conditions.  For the case of the model considered here, the solutions that describe vector mesons in the vacuum are obtained solving eq. (\ref{eqmotion}) with the condition of vanishing at the boundary  $ V (p, z=0) =0$
 that selects the normalizable ones.  Previous works of quasinormal modes of mesons in the softwall model can be found, for example,  in \cite{Miranda:2009uw,Mamani:2013ssa,Mamani:2018uxf}. An alternative approach to study the thermal masses of  mesons and baryons in the softwall model at low temperature can be found in  \cite{Gutsche:2019blp,Gutsche:2019pls}.

 At finite temperature, the objects that play the equivalent role are quasinormal modes, that represent  quasi-states.  They are field solutions satisfying, in the black hole geometry, an incoming wave condition at the horizon, and, as in the zero temperature case, a Dirichlet condition $ V (p, z=0) =0$ on the boundary. 
These solutions exist for complex frequencies $\omega=\omega_{R}  + i \omega_{I}$ with an important physical interpretation for the frequency components: the real part is related to the thermal mass of the vector mesons while the imaginary part is related to the dissociation degree of the  quasi-particle states.

\begin{figure}[h]
\label{g67}
\begin{center}
\includegraphics[scale=0.35]{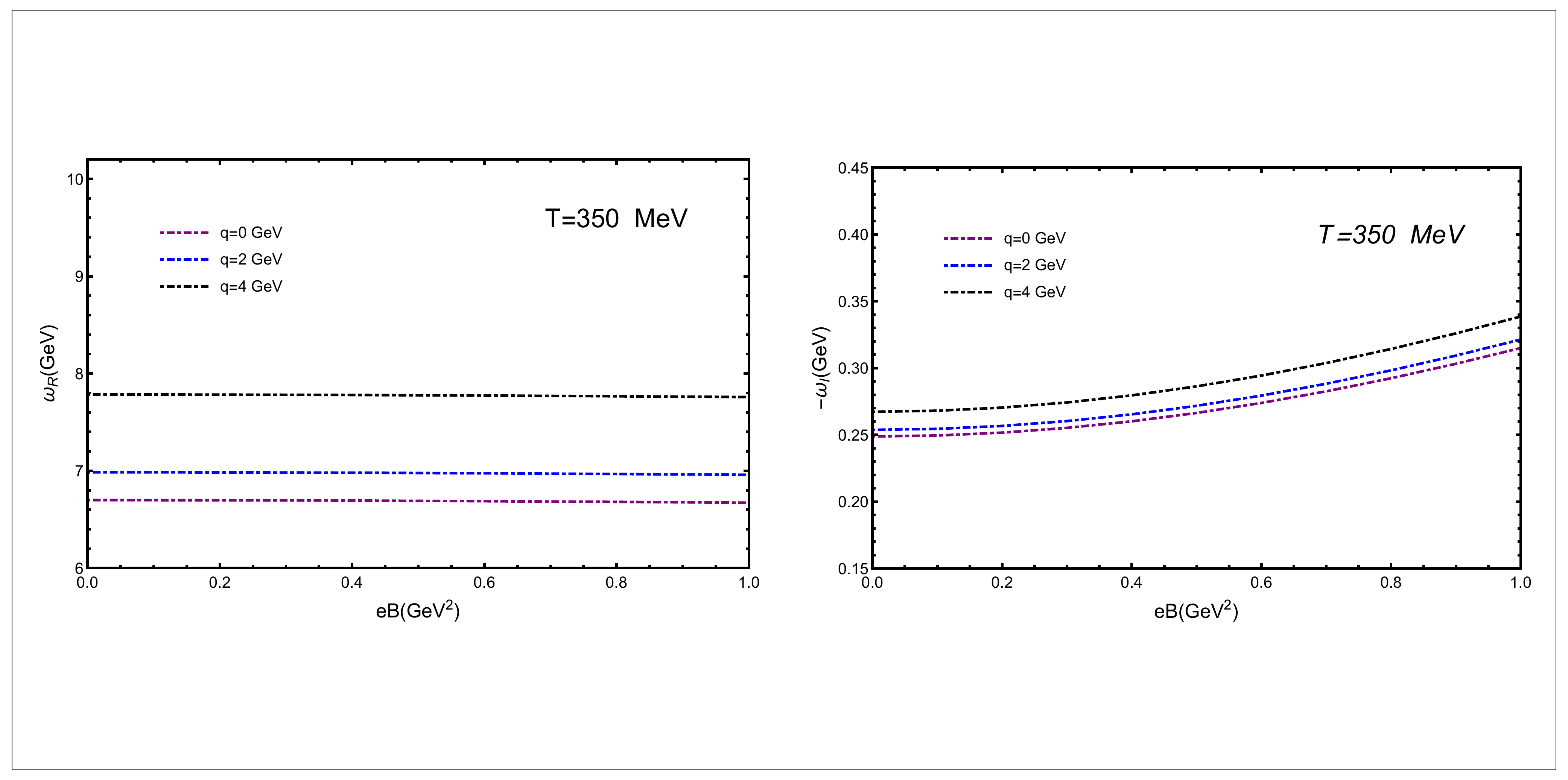}
\end{center}
\caption{Real (left panel) and imaginary (right panel) parts of the quasinormal mode frequency of bottomonium with transverse polarization  at T = 350 MeV  as a function of the magnetic field for different values of the linear momentum.      }
\end{figure}

\begin{figure}[h]
\label{g67}
\begin{center}
\includegraphics[scale=0.35]{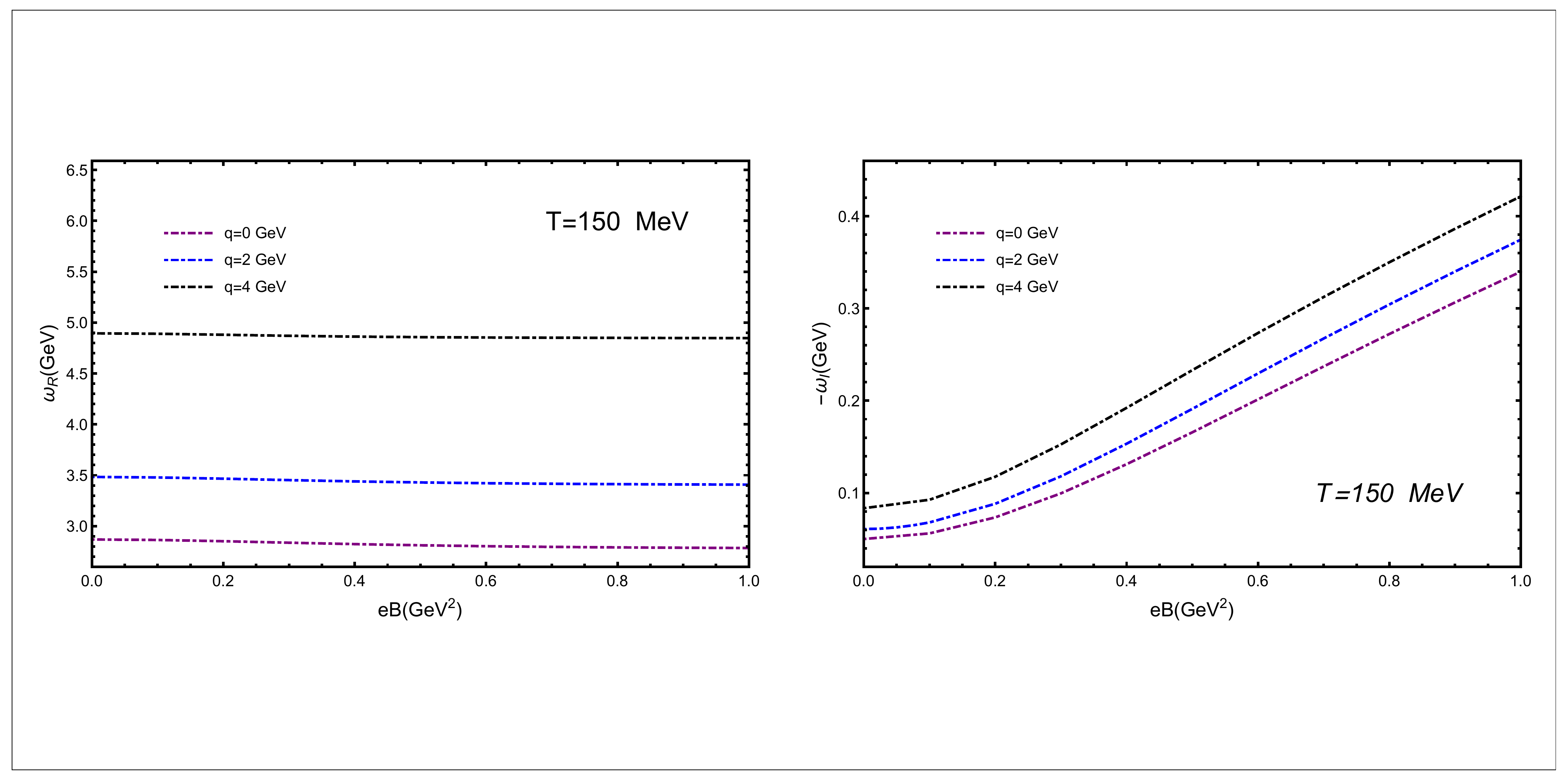}
\end{center}
\caption{Real (left panel) and imaginary (right panel) parts of the quasinormal mode frequency of charmonium with transverse polarization at T = 150 MeV  as a function of the magnetic field for different values of the linear momentum.          }
\end{figure}

\subsection{ Equations, boundary conditions and numerical approach} 

The quasinormal modes can be numerically evaluated from the equations of motion (\ref{eqTransx}) and (\ref{eqLogx}) using the shooting method of refs. \cite{Kaminski:2008ai,Kaminski:2009ce,Amado:2009ts,Kaminski:2009dh,Janiszewski:2015ura}. One needs to specify two boundary conditions at the horizon and then finds the correct parameter, given by the complex frequency, which satisfies the normalization condition of the field on the boundary. Therefore,  equations (\ref{eqTransx}) and (\ref{eqLogx}  need to be solved
 \begin{equation}\label{eqTransx2}
\mathcal{E}_{\alpha}''+\left(\frac{f'}{f}-\frac{1}{z}-\phi'+\frac{h'}{2h} \right)\mathcal{E}_{\alpha}'+\left(\frac{\omega^2}{f^2}-\frac{q^2}{fh}\right)\mathcal{E}_{\alpha}=0 \, , \,\,\,\,\,\,\,\,\,\,\ (\alpha=1,2) 
\end{equation}

\begin{equation}\label{eqLogx2}
\mathcal{E}_{3}''+\left(\frac{f'}{f}\frac{\omega^2}{\omega^2 h-q^2f}-\frac{1}{z}-\phi' +\frac{d'}{d}-\frac{h'}{2h}\frac{\omega^2 h+q^2f}{\omega^2 h-q^2f} \right)\mathcal{E}_{3}'+\left(\frac{\omega^2}{f^2}-\frac{q^2}{fh}\right)\mathcal{E}_{3}=0. \,\,\,\,\,\,\,\,\,\,\ 
\end{equation} 
using the infalling boundary conditions at the horizon, presented in eq. (\ref{boundary}),
\begin{eqnarray}\label{infa1}
\lim_{z\rightarrow z_h}\mathcal{E}_{j}(z,p)&=& \left(1-\frac{z}{z_h}\right)^{-i\omega/4 \pi T}\,\left[1+a_{1}\left(1-\frac{z}{z_h}\right)+\cdots\right],
\end{eqnarray}
and the derivative of the infalling  condition : 
\begin{eqnarray}\label{infa2}
 \lim_{z\rightarrow z_h} \partial_{z} \mathcal{E}_{j}(z,p)&=& \left(1-\frac{z}{z_h}\right)^{-i\omega/4 \pi T}\,\left[\frac{-a_{1}}{z_{h}}+\frac{-2a_{2}}{z_{h}}\left( 1- \frac{z}{z_h}\right)\cdots\right] \cr  &-& \frac{i\omega}{(4 \pi T)}\left(1-\frac{z}{z_h}\right)^{-1}\lim_{z\rightarrow z_h}\mathcal{E}_{j}(z,p).
\end{eqnarray}
The coefficients that appear in the infalling condition can be determined inserting this condition into the equations of motion. Similarly to normal modes, the quasinormal modes exist only for a discrete set of frequencies $\omega_{n}$, but in this case they are complex.

\begin{figure}[h]
\label{g67}
\begin{center}
\includegraphics[scale=0.35]{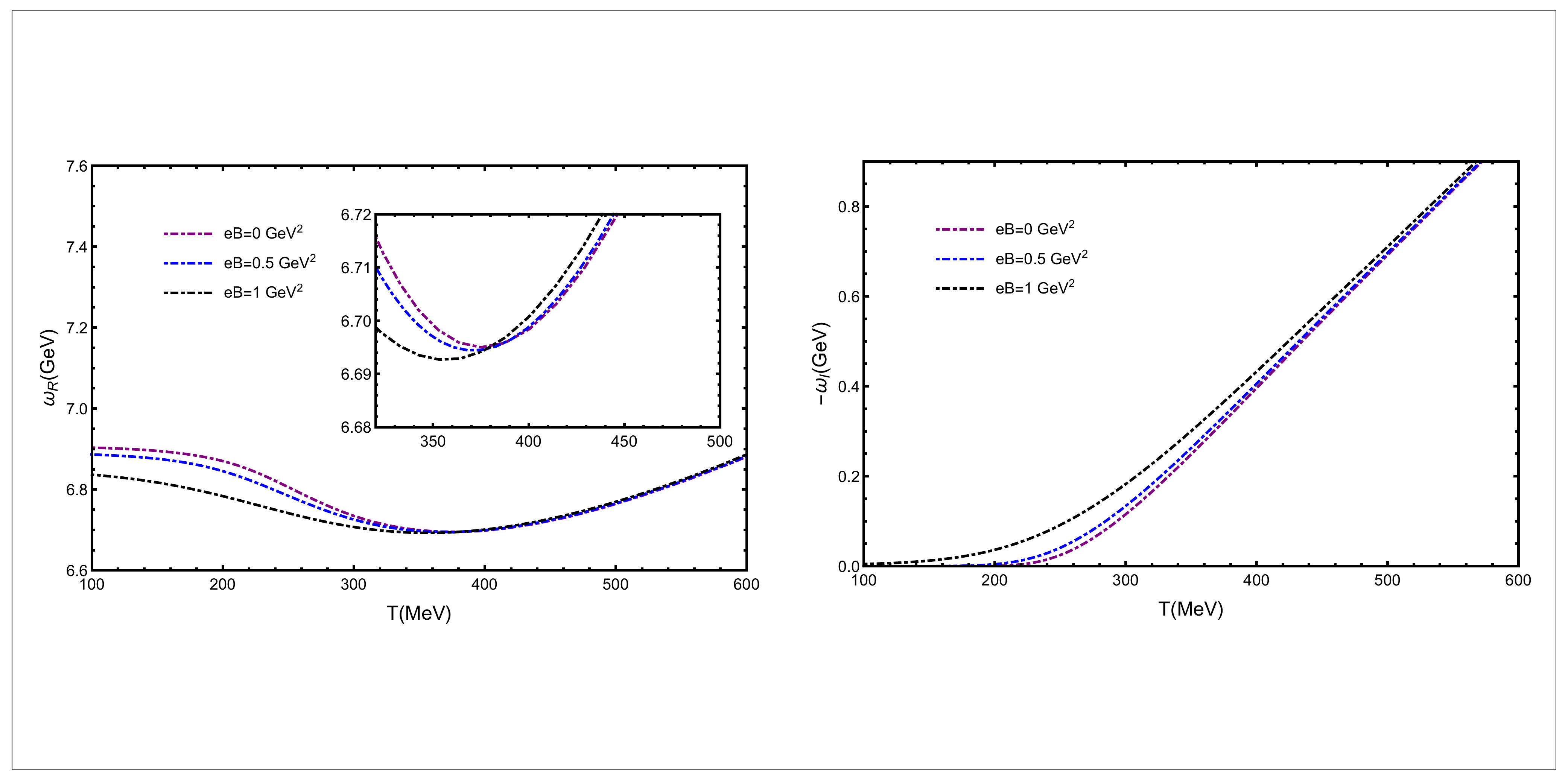}
\end{center}
\caption{Real (left panel) and imaginary (right panel) parts of the quasinormal mode frequency of bottomonium with longitudinal  polarization as a function of the temperature for different values of magnetic field.  In the detail:  behaviour of the  real part near 400 MeV.   }
\end{figure}

\begin{figure}[h]
\label{g67}
\begin{center}
\includegraphics[scale=0.35]{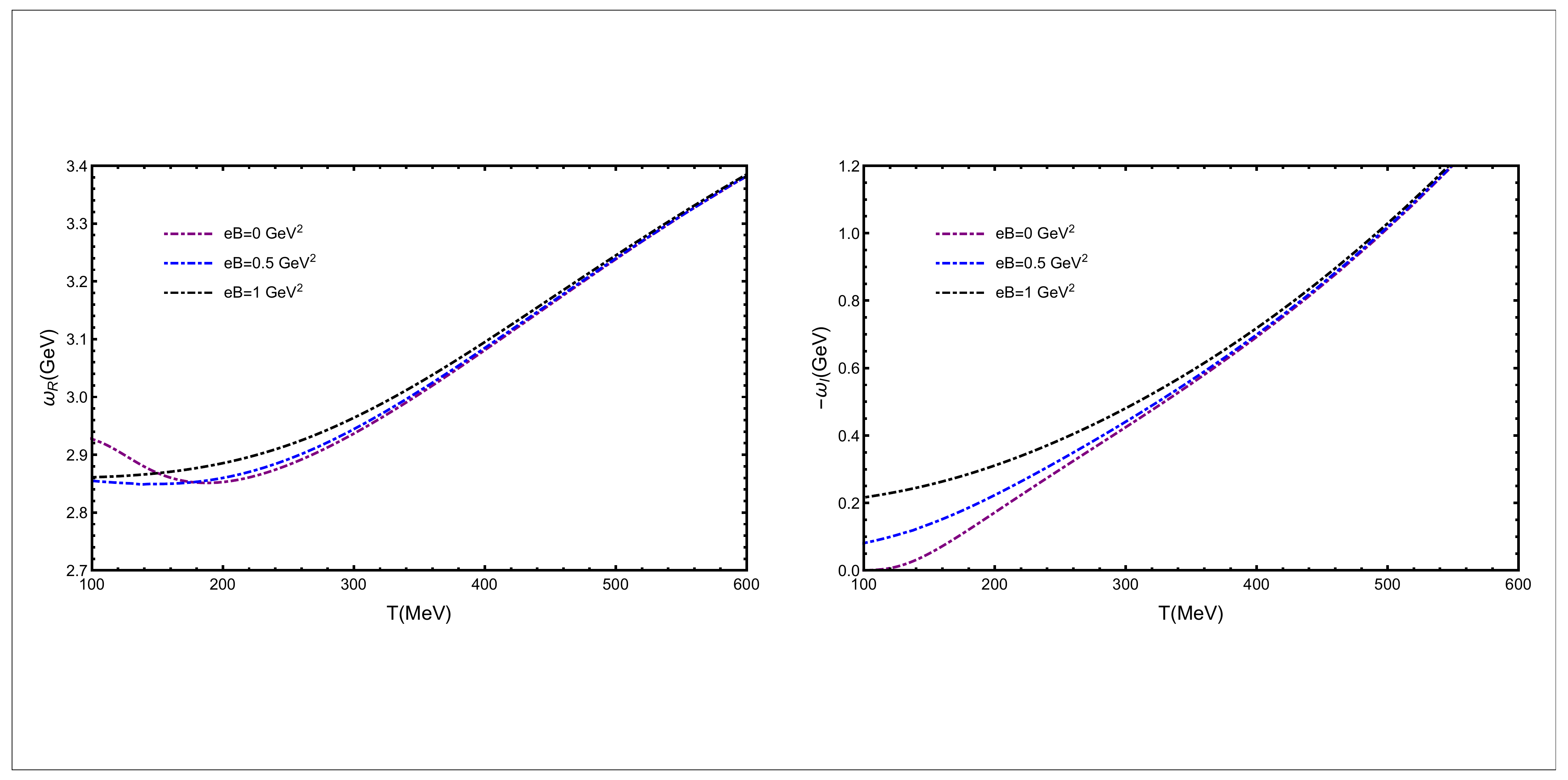}
\end{center}
\caption{Real (left panel) and imaginary (right panel) parts of the quasinormal mode frequency of charmonium with longitudinal  polarization as a function of the temperature for different values of magnetic field.       }
\end{figure}

\subsection{ Exploring the results } 

The complex frequencies, that we solve numerically, depend on the temperature $T$ , magnetic field $eB$, linear momentum $q $ and on the flavour: bottomonium or charmonium. We choose,  for each flavour, combinations of the other parameters that give a representative picture. The dissociation process corresponds to an increase in the imaginary part of the frequency, while the variation of the real part represents a change in the thermal mass. 

Let us start with vector mesons that have a polarization transverse to the direction of the magnetic field. 
We show in figures {\bf 3} and {\bf 4}   the complex quasinormal frequencies as functions of the temperature, for bottomonium and charmonium, respectively,  for three different values of the magnetic field background $eB$. One notes clearly an increase in the imaginary part of the frequency with the value of $eB$. This means that the presence of the magnetic field enhances the dissociation of the mesons by the thermal medium. This effect is more noticeable at lower temperatures where the dissociation by thermal effect is not so strong,  due the fact that we considered only the indirect effect of the magnetic field caused by a modification in the thermal medium. On the other hand, the real part, that represents the thermal mass decreases with $eB$ for low temperatures. This effect is not noticeable for temperatures above 300 MeV. 

\begin{figure}[h]
\label{g67}
\begin{center}
\includegraphics[scale=0.35]{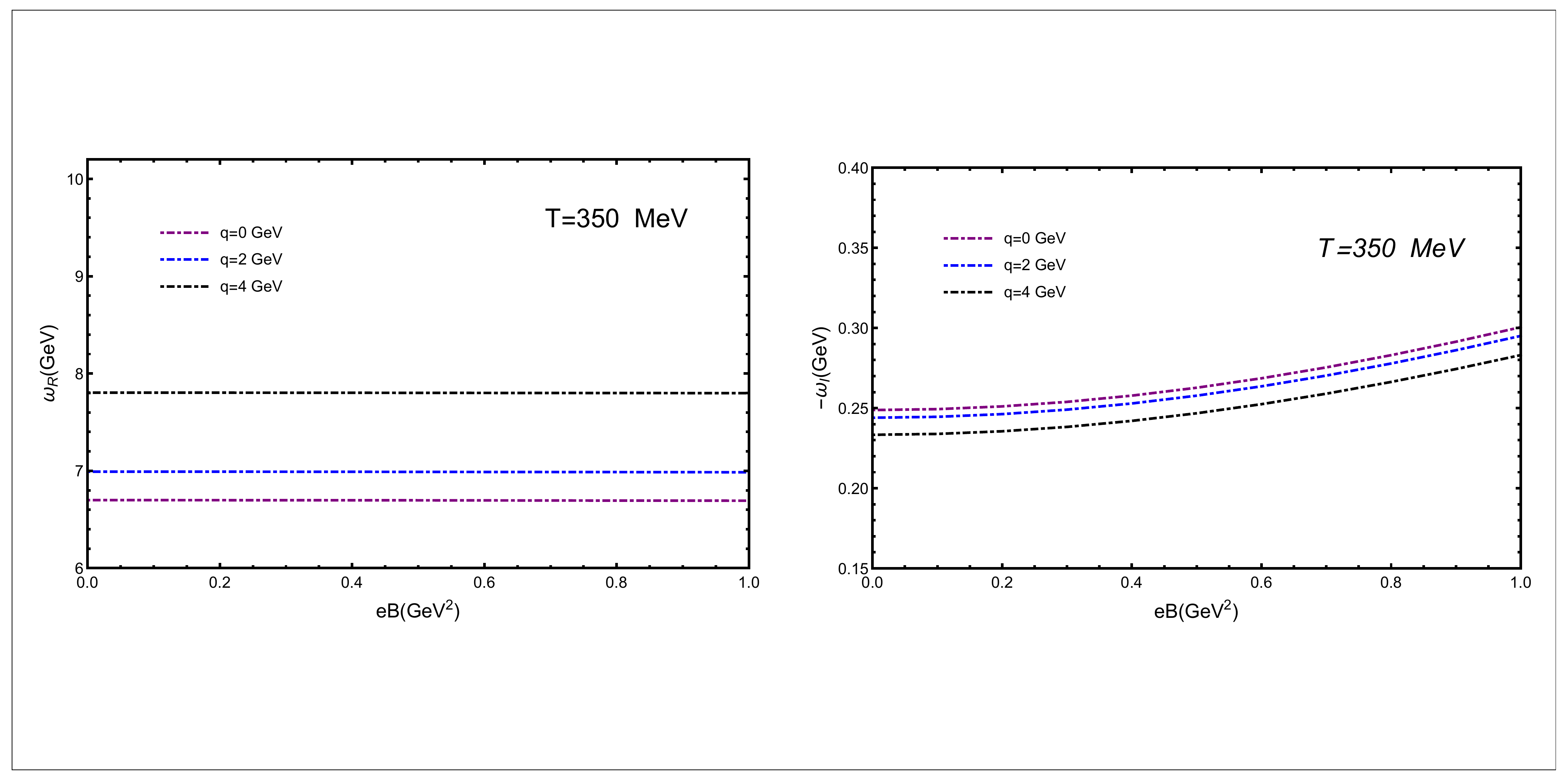}
\end{center}
\caption{Real (left panel) and imaginary (right panel) parts of the quasinormal mode frequency of bottomonium with longitudinal  polarization  at T = 350 MeV  as a function of the magnetic field for different values of the linear momentum.      }
\end{figure}

\begin{figure}[h]
\label{g67}
\begin{center}
\includegraphics[scale=0.35]{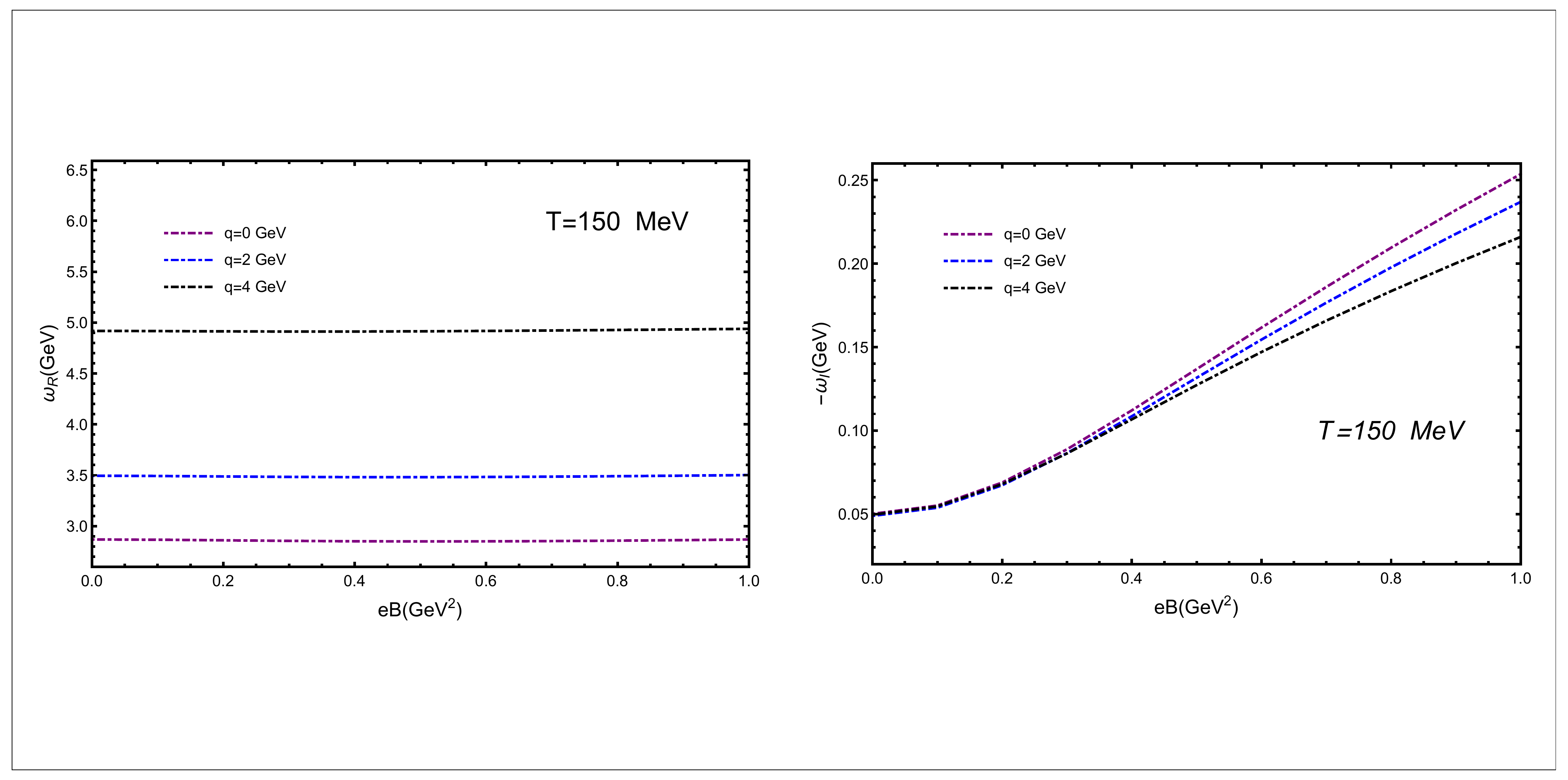}
\end{center}
\caption{Real (left panel) and imaginary (right panel) parts of the quasinormal mode frequency of charmonium with longitudinal polarization at T = 150 MeV  as a function of the magnetic field for different values of the linear momentum.   }
\end{figure}

Then, in figures {\bf 5} and {\bf 6},  we analyse the effect of motion of heavy mesons relative to the magnetized medium, when the polarization is transverse to the motion and to the $eB$ field. We plot the quasinormal frequencies for bottomonium at T = 350 MeV and  charmonium at T = 150 MeV, respectively, for three different values of linear momentum,  as a function of the magnetic field. One can see that 
the magnetic field produces an increase in the imaginary part of the frequency, associated with the dissociation degree. On the other hand, the motion of the meson produces an  increase in the thermal mass and also in the imaginary part of the frequency. 

Then we move to the longitudinal case, when the polarization is parallel to the direction of the magnetic field. We show in figures  {\bf 7} and {\bf 8}    the quasinormal frequencies as functions of the temperature, for bottomonium and charmonium, respectively,  for three different values of the magnetic field background $eB$. The imaginary part of the frequency increases with the value of $eB$, as it happens in the transverse case. So, the presence of the magnetic field enhances the dissociation of the mesons. For the real part, that represents the thermal mass, at low temperatures magnetic field produces a decrease. However, for charmonium at about 200 MeV and bottomonium at about 380 MeV the behaviour changes. It should be notes however that charmonium at T= 200 MeV and bottomonium at T = 380 MeV are in a very high degree of dissociation. So, the quasi particle state is highly mixed with the thermal medium.

 Finally, in  figures {\bf 9} and {\bf 10},  we show the spectral functions of heavy quarks in motion  relative to the magnetized medium, when the polarization is parallel  to the motion and to the $eB$ field. We plot the quasinormal frequencies at the representative temperatures of  T = 350 MeV for bottomonium  and  T = 150 MeV for charmonium, for three different values of linear momentum,  as a function of the magnetic field. 
 One can see that the magnetic field produces an increase in the imaginary part of the frequency.   On the other hand,  motion of the mesons, relative to the plasma,  has the opposite effect: the higher the momentum, the smaller the imaginary part. So, in this longitudinal case, mesons in motion have a slower dissociation degree.  For the real part of the frequency, the motion of the meson produces an  increase.

\section{Conclusions}
 
 We presented the calculation of the complex quasinormal mode frequencies for heavy vector mesons in a plasma with a constant and uniform background magnetic field. The picture that emerged is consistent with previous analysis using the spectral function approach. The magnetic field produces an increase in the imaginary part of the frequencies, that corresponds to the broadening of the quasiparticle peaks in the spectral function approach. Physically this corresponds to an increase in the dissociation degree with the magnetic field intensity. 
 
 We considered only the   effect of the magnetic field  in the thermal medium,  encoded in the background metric of eq.(\ref{metric2}). It is important to note that it is possible to study the direct effect of the field on the charged constituents of the meson following the approach of  \cite{Dudal:2014jfa,Dudal:2018rki}.   An interesting topic for future investigation would be to calculate the quasinormal model following this alternative approach, that uses a DBI action.

\noindent {\bf Acknowledgments:}   N.B. is partially supported by CNPq and L. F. Ferreira is supported by CNPq under Grant No. 153337/2018-4.


\begin{thebibliography}{ABC}


\bibitem{Matsui:1986dk} 
  T.~Matsui and H.~Satz,
  Phys.\ Lett.\ B {\bf 178}, 416 (1986).
 doi:10.1016/0370-2693(86)91404-8.
 
 \bibitem{Satz:2005hx} 
  H.~Satz,
  J.\ Phys.\ G {\bf 32}, R25 (2006)
  doi:10.1088/0954-3899/32/3/R01
  [hep-ph/0512217].
 
\bibitem{Braga:2017bml} 
  N.~R.~F.~Braga, L.~F.~Ferreira and A.~Vega,
  Phys.\ Lett.\ B {\bf 774}, 476 (2017)
  [arXiv:1709.05326 [hep-ph]]. 

  
 
\bibitem{Braga:2018zlu} 
  N.~R.~F.~Braga and L.~F.~Ferreira,
  Phys.\ Lett.\ B {\bf 783}, 186 (2018)
  [arXiv:1802.02084 [hep-ph]].
 
\bibitem{Braga:2018hjt} 
  N.~R.~F.~Braga and L.~F.~Ferreira,
  JHEP {\bf 1901}, 082 (2019)
  doi:10.1007/JHEP01(2019)082
  [arXiv:1810.11872 [hep-ph]].
 
 
 
 
 
 
 
\bibitem{Braga:2015jca} 
  N.~R.~F.~Braga, M.~A.~Martin Contreras and S.~Diles,
  Phys.\ Lett.\ B {\bf 763}, 203 (2016)
  [arXiv:1507.04708 [hep-th]].
 
 \bibitem{Braga:2016wkm} 
  N.~R.~F.~Braga, M.~A.~Martin Contreras and S.~Diles,
  Eur.\ Phys.\ J.\ C {\bf 76}, no. 11, 598 (2016)
  [arXiv:1604.08296 [hep-ph]].
      
  
 
\bibitem{Braga:2017oqw} 
  N.~R.~F.~Braga and L.~F.~Ferreira,
  Phys.\ Lett.\ B {\bf 773}, 313 (2017)
  [arXiv:1704.05038 [hep-ph]].
 
 
 
\bibitem{Kharzeev:2007jp} 
  D.~E.~Kharzeev, L.~D.~McLerran and H.~J.~Warringa,
  Nucl.\ Phys.\ A {\bf 803}, 227 (2008)
  doi:10.1016/j.nuclphysa.2008.02.298
  [arXiv:0711.0950 [hep-ph]].
  
\bibitem{Fukushima:2008xe} 
  K.~Fukushima, D.~E.~Kharzeev and H.~J.~Warringa,
  Phys.\ Rev.\ D {\bf 78}, 074033 (2008)
  doi:10.1103/PhysRevD.78.074033
  [arXiv:0808.3382 [hep-ph]].
    
\bibitem{Skokov:2009qp} 
  V.~Skokov, A.~Y.~Illarionov and V.~Toneev,
  Int.\ J.\ Mod.\ Phys.\ A {\bf 24}, 5925 (2009)
  doi:10.1142/S0217751X09047570
  [arXiv:0907.1396 [nucl-th]].
    
    
    
    
    
    
    
\bibitem{Bali:2011qj} 
  G.~S.~Bali, F.~Bruckmann, G.~Endrodi, Z.~Fodor, S.~D.~Katz, S.~Krieg, A.~Schafer and K.~K.~Szabo,
  JHEP {\bf 1202}, 044 (2012)
  doi:10.1007/JHEP02(2012)044
  [arXiv:1111.4956 [hep-lat]].
    
    
\bibitem{Fraga:2012fs} 
  E.~S.~Fraga and L.~F.~Palhares,
  Phys.\ Rev.\ D {\bf 86}, 016008 (2012)
  doi:10.1103/PhysRevD.86.016008
  [arXiv:1201.5881 [hep-ph]].
    
    
\bibitem{Ballon-Bayona:2013cta} 
  A.~Ballon-Bayona,
  JHEP {\bf 1311}, 168 (2013)
  doi:10.1007/JHEP11(2013)168
  [arXiv:1307.6498 [hep-th]].
    
\bibitem{Mamo:2015dea} 
  K.~A.~Mamo,
  JHEP {\bf 1505}, 121 (2015)
  doi:10.1007/JHEP05(2015)121
  [arXiv:1501.03262 [hep-th]].  
    
\bibitem{Dudal:2015wfn} 
  D.~Dudal, D.~R.~Granado and T.~G.~Mertens,
  Phys.\ Rev.\ D {\bf 93}, no. 12, 125004 (2016)
  doi:10.1103/PhysRevD.93.125004
  [arXiv:1511.04042 [hep-th]].
    
    \bibitem{Evans:2016jzo} 
  N.~Evans, C.~Miller and M.~Scott,
  Phys.\ Rev.\ D {\bf 94}, no. 7, 074034 (2016)
  doi:10.1103/PhysRevD.94.074034
  [arXiv:1604.06307 [hep-ph]].
    
\bibitem{Li:2016gfn} 
  D.~Li, M.~Huang, Y.~Yang and P.~H.~Yuan,
  JHEP {\bf 1702}, 030 (2017)
  doi:10.1007/JHEP02(2017)030
  [arXiv:1610.04618 [hep-th]].
    
  \bibitem{Ballon-Bayona:2017dvv} 
  A.~Ballon-Bayona, M.~Ihl, J.~P.~Shock and D.~Zoakos,
  JHEP {\bf 1710}, 038 (2017)
  doi:10.1007/JHEP10(2017)038
  [arXiv:1706.05977 [hep-th]].
    
    
\bibitem{Rodrigues:2017cha} 
  D.~M.~Rodrigues, E.~Folco Capossoli and H.~Boschi-Filho,
  arXiv:1709.09258 [hep-th].  
     
     

 
 
\bibitem{Iwasaki:2018pby} 
  S.~Iwasaki, M.~Oka, K.~Suzuki and T.~Yoshida,
  Phys.\ Lett.\ B {\bf 790}, 71 (2019)
  doi:10.1016/j.physletb.2018.10.024
  [arXiv:1802.04971 [hep-ph]].
\bibitem{Giataganas:2018uuw} 
  D.~Giataganas,
  Phys.\ Rev.\ D {\bf 98}, no. 10, 106010 (2018)
  doi:10.1103/PhysRevD.98.106010
  [arXiv:1805.08245 [hep-th]].


\bibitem{Iwasaki:2018czv} 
  S.~Iwasaki and K.~Suzuki,
  Phys.\ Rev.\ D {\bf 98}, no. 5, 054017 (2018)
  doi:10.1103/PhysRevD.98.054017
  [arXiv:1805.09787 [hep-ph]].
  
  
\bibitem{Bonati:2018uwh} 
  C.~Bonati, S.~Calì, M.~D'Elia, M.~Mesiti, F.~Negro, A.~Rucci and F.~Sanfilippo,
  Phys.\ Rev.\ D {\bf 98}, no. 5, 054501 (2018)
  doi:10.1103/PhysRevD.98.054501
  [arXiv:1807.01673 [hep-lat]].

      
      
    
\bibitem{Polchinski:2001tt}
  J.~Polchinski and M.~J.~Strassler,
  Phys.\ Rev.\ Lett.\  {\bf 88}, 031601 (2002)
  [arXiv:hep-th/0109174].

\bibitem{BoschiFilho:2002ta}
  H.~Boschi-Filho and N.~R.~F.~Braga,
  Eur.\ Phys.\ J.\  C {\bf 32}, 529 (2004)
  [arXiv:hep-th/0209080].
  
\bibitem{BoschiFilho:2002vd}
  H.~Boschi-Filho and N.~R.~F.~Braga,
  JHEP {\bf 0305}, 009 (2003)
  [arXiv:hep-th/0212207].
   
    
    
    
    
    
    
    
  
  
  
  
  
  
  
  
  
  
   
  
\bibitem{Son:2002sd} 
  D.~T.~Son and A.~O.~Starinets,
  JHEP {\bf 0209}, 042 (2002)
  doi:10.1088/1126-6708/2002/09/042
  [hep-th/0205051].


\bibitem{Mamani:2013ssa} 
  L.~A.~H.~Mamani, A.~S.~Miranda, H.~Boschi-Filho and N.~R.~F.~Braga,
  JHEP {\bf 1403}, 058 (2014)
  doi:10.1007/JHEP03(2014)058
  [arXiv:1312.3815 [hep-th]].
  
  
\bibitem{Miranda:2009uw} 
  A.~S.~Miranda, C.~A.~Ballon Bayona, H.~Boschi-Filho and N.~R.~F.~Braga,
  JHEP {\bf 0911}, 119 (2009)
  doi:10.1088/1126-6708/2009/11/119
  [arXiv:0909.1790 [hep-th]].
  
\bibitem{Mamani:2018uxf} 
  L.~A.~H.~Mamani, A.~S.~Miranda and V.~T.~Zanchin,
  arXiv:1809.03508 [hep-th].



\bibitem{Gutsche:2019blp} 
  T.~Gutsche, V.~E.~Lyubovitskij, I.~Schmidt and A.~Y.~Trifonov,
  Phys.\ Rev.\ D {\bf 99}, no. 5, 054030 (2019)
  doi:10.1103/PhysRevD.99.054030
  [arXiv:1902.01312 [hep-ph]].

\bibitem{Gutsche:2019pls} 
  T.~Gutsche, V.~E.~Lyubovitskij, I.~Schmidt and A.~Y.~Trifonov,
  arXiv:1905.02577 [hep-ph].

\bibitem{Kaminski:2008ai} 
  M.~Kaminski,
  Fortsch.\ Phys.\  {\bf 57}, 3 (2009)
  [arXiv:0808.1114 [hep-th]].	
	
\bibitem{Kaminski:2009ce} 
  M.~Kaminski, K.~Landsteiner, F.~Pena-Benitez, J.~Erdmenger, C.~Greubel and P.~Kerner,
  JHEP {\bf 1003}, 117 (2010)
  [arXiv:0911.3544 [hep-th]].
  
  
\bibitem{Amado:2009ts} 
  I.~Amado, M.~Kaminski and K.~Landsteiner,
  JHEP {\bf 0905}, 021 (2009)
  doi:10.1088/1126-6708/2009/05/021
  [arXiv:0903.2209 [hep-th]].
  
\bibitem{Kaminski:2009dh} 
  M.~Kaminski, K.~Landsteiner, J.~Mas, J.~P.~Shock and J.~Tarrio,
  JHEP {\bf 1002}, 021 (2010)
  doi:10.1007/JHEP02(2010)021
  [arXiv:0911.3610 [hep-th]].
  
    
     
    
    
    
    
\bibitem{Janiszewski:2015ura} 
  S.~Janiszewski and M.~Kaminski,
  Phys.\ Rev.\ D {\bf 93}, no. 2, 025006 (2016)
  [arXiv:1508.06993 [hep-th]].
  
   

	

  
\bibitem{Dudal:2014jfa} 
  D.~Dudal and T.~G.~Mertens,
  Phys.\ Rev.\ D {\bf 91}, 086002 (2015)
  doi:10.1103/PhysRevD.91.086002
  [arXiv:1410.3297 [hep-th]].
 
\bibitem{Dudal:2018rki} 
  D.~Dudal and T.~G.~Mertens,
  Phys.\ Rev.\ D {\bf 97}, no. 5, 054035 (2018)
  doi:10.1103/PhysRevD.97.054035
  [arXiv:1802.02805 [hep-th]].
 
 
 

\end{thebibliography}
 \end{document}